\documentstyle[aps,preprint]{revtex}
\tightenlines 
\begin{document}
\draft
\title{First Passage Time Distribution for Anomalous Diffusion}
\author{Govindan Rangarajan$^{1,*}$ and Mingzhou Ding$^{2,\dagger}$}
\address{$^1$Department of Mathematics and Center for Theoretical
Studies, Indian Institute of Science, Bangalore 560 012, India}
\address{$^2$Center for Complex Systems and Brain Sciences,
Florida Atlantic University, Boca Raton, FL 33431, USA}
\maketitle

\begin{abstract}
{We study the first passage time (FPT) problem in Levy type of
anomalous diffusion. Using the recently formulated fractional
Fokker-Planck equation, we obtain an analytic expression for the
FPT distribution which, in the large passage time limit, is
characterized by a universal power law. Contrasting this power law
with the asymptotic FPT distribution from another type of
anomalous diffusion exemplified by the fractional Brownian motion,
we show that the two types of anomalous diffusions give rise to
two distinct scaling behavior.}

\end{abstract}
\newpage
\section{Introduction}

For a stochastic process, the first passage time (FPT) is defined
as the time $T$ when the process, starting from a given point,
reaches a predetermined level for the first time, and is a random
variable \cite{risken}. Escape times from a random potential,
intervals between neural spikes, and fatigue failure times of
engineering structures are all examples of FPTs, arising in
physics \cite{gardiner}, biology \cite{tuckwell}, and engineering
\cite{cai}, respectively. Thus, knowledge of the FPT distribution,
$f(t)$, is essential for the effective application of
probabilistic analysis. (As a convention we use capital letters to
denote random variables and lower case letters to denote their
values.) Unfortunately, only in very few cases does one have
explicit analytical expressions for $f(t)$. One such case is the
ordinary Brownian motion, an example of ordinary diffusion, in
which the FPT is described by the famous inverse Gaussian law
\cite{grimmet}. The main contribution of this work is the
derivation of the exact solution of $f(t)$ for a much broader
class of stochastic processes, namely, the Levy type of anomalous
diffusion \cite{shlesinger,list} in which the mean square
displacement of the diffusive variable $X(t)$ scales with time as
$<X^2(t)> \sim t^{\gamma}$ with $0<\gamma<2$. Specifically, using
a recently formulated framework of fractional Fokker-Planck
equation (FFPE) \cite{metzler}, we express $f(t)$ in terms of Fox
or H-functions \cite{fox,mathai}, which is shown to contain the
inverse Gaussian distribution as a special case. Furthermore, we
show that in the asymptotic limit of large $t$, $f(t)$ scales with
$t$ as $f(t) \sim t^{-1-\gamma/2}$. Our next result concerns the
comparison with a different type of anomalous diffusion,
represented by fractional Brownian motion (fBm)
\cite{mandelbrot,taqqu}, where again $<X^2(t)> \sim t^{\gamma}$
with $0<\gamma<2$. For this type, we argue that $f(t) \sim
t^{\gamma/2-2}$ for large $t$, a result that has been conjectured
earlier \cite{ding}. Finally, we present numerical simulations
which verify the analytical results.

\section{FFPE and Derivation of FPT Distribution}

The Levy type of anomalous diffusion considered in this work is a
class of non Gaussian and non Markovian processes founded on the
continuous time random walk (CTRW) where the waiting time obeys
certain power law distribution \cite{shlesinger}. Let $\phi(y,u)$
denote the joint probability density between the waiting time $U$
and the jump size $Y$. It can be shown that, depending on the
specific form of $\phi(y,u)$, the CTRW can produce both
subdiffusive ($0 < \gamma < 1$) and superdiffusive processes ($1 <
\gamma < 2$) as well as ordinary diffusion ($\gamma=1$)
\cite{shlesinger,wang}. For example, consider
\begin{equation}\label{LWpdf}
\phi(y,u) = \frac{1}{\sqrt{2 \pi \sigma^2}} \exp[-y^2/2 \sigma^2]
\frac{(\alpha-1)/\tau}{(1+u/\tau)^{\alpha}},
\end{equation}
where $Y$ and $U$ are decoupled with $Y$ being a Gaussian
variable. (We note that, strictly speaking, the distribution of
$U$ is not a Levy stable distribution, but belongs to the domain
of attraction \cite{taqqu} of a one-sided stable Levy law. We call
$U$ a ``Levy type of variable'' for want of a better name.) For $1
< \alpha < 2$, the corresponding CTRW is characterized by a
subdiffusive process with $\gamma=\alpha-1$, and for $\alpha \ge
2$, one gets ordinary diffusion with $\gamma=1$. If, on the other
hand, $Y$ and $U$ are coupled through
\begin{equation}\label{LW2pdf}
\phi(y,u) = \frac{1}{2} \delta(u/\tau-|y|/\sigma)
\frac{(\beta-1)/\tau}{(1+u/\tau)^{\beta}},
\end{equation}
where $2 < \beta < 3$ and $\delta(\cdot)$ is the Dirac delta
function, the CTRW describes a superdiffusive process with
$\gamma=4-\beta$.

Let $W(x,t)$ be the probability density function for a CTRW $X(t)$
with $X(0)=0$. Consider the generalized diffusion limit where
$\sigma$ and $\tau$ are scaling parameters for the space and time
variables. For the subdiffusive case, this means taking the limit
$\sigma^2 \to 0$ and $\tau \to 0$ with $K=\sigma^2/2
\Gamma(1-\gamma) \tau^{\gamma}$ kept a constant, and for the
superdiffusive process, this means taking the same limit with $K
=(3-\gamma)(2-\gamma)\Gamma(\gamma -1)\sigma^2/ 2 (5-2\gamma)
\tau^{\gamma}$ kept a constant. In this limit it can be shown
\cite{shlesinger} that the evolution of $W(x,t)$ is determined by
the following FFPE \cite{metzler}:
\begin{equation}
W(x,t)-W(x,0) =\  _0 D_t^{-\gamma} K \frac{\partial^2}{\partial
x^2} W(x,t), \ \ \ 0 < \gamma < 2, \label{FFPE}
\end{equation}
where the Riemann-Liouville fractional integral operator $_0
D_t^{-\gamma}$ is defined as \cite{oldham,miller}
\begin{equation}
_0 D_t^{-\gamma} W(x,t) = \frac{1}{\Gamma(\gamma)} \int_0^t \ dt'
\ (t-t')^{\gamma-1}W(x,t'), \ \ \ \gamma > 0,
\end{equation}
with $\Gamma(z)$ being the gamma function \cite{gradshteyn}. The
constant $K$ is the generalized diffusion constant defined in the
above generalized diffusion limit. From Eq.~(\ref{FFPE}) it is
easily shown that $<X^2(t)>=2Kt^{\gamma}/\Gamma(1+\gamma)$.

In the framework of the Fokker-Planck equation, the first passage
time problem is defined in terms of having absorbing boundaries at
$x=-\infty$ and $x=a$, where $a$ is the predetermined level of
crossing, with the initial condition $W(x,0)=\delta(x)$
\cite{risken}. An equivalent formulation, due to symmetry, is to
solve Eq.~(\ref{FFPE}) with the following boundary and initial
conditions: $W(0,t)=0$, $W(\infty,t) = 0$, $W(x,0) =\delta(x-a)$,
where $x=a$ is the new starting point of the process, containing
the initial concentration of the distribution. (This latter
formulation makes the subsequent derivation less cumbersome.) Once
we solve for $W(x,t)$, the first passage time distribution $f(T)$
is given by \cite{risken}
\begin{equation}
f(t) = - \frac{d}{dt} \int_0^{\infty}\ dx \ W(x,t).
\label{fT}\end{equation}

Taking into account of the boundary and initial conditions we are
led to the following expansion for $W(x,t)$ \cite{morse}
\begin{equation}
W(x,t) = \frac{2}{\pi} \int_0^{\infty}\  dk \sin kx \ \sin ka \
A(k,t), \label{sol1}\end{equation} with $A(k,0)=1$. To determine
the unknown function $A(k,t)$, we substitute the above expansion
for $W(x,t)$ in Eq.~(\ref{FFPE}) and, after straightforward
algebra, obtain $A(k,t)-1=-K k^2\ _0 D_t^{-\gamma} A(k,t)$. Taking
the Laplace transform with respect to $t$, we have
\begin{equation}
A(k,p) = \frac{1}{p+k^2 K p^{1-\gamma}}, \label{lap}
\end{equation}
where $A(k,p)$ is the Laplace transform of $A(k,t)$. Here we have
applied the result \cite{miller} that the Laplace transform of $_0
D_t^{-\gamma} A(k,t)$ is $A(k,p)/p^{\gamma}$. Inverse Laplace
transform of Eq.~(\ref{lap}) yields \cite{erdelyi}
\begin{equation}
A(k,t) = E_{\gamma}(-k^2 K t^{\gamma}), \label{akt}
\end{equation}
where $E_{\gamma}(z)$ is the Mittag-Leffler function
\cite{erdelyi}. Substituting Eq.~(\ref{akt}) into Eq. (\ref{sol1})
we get
\begin{equation}
W(x,t) = \frac{2}{\pi} \int_0^{\infty}\  dk \sin kx \ \sin ka \
E_{\gamma}(-k^2 K t^{\gamma}). \label{sol2}\end{equation}

To proceed further, we introduce the Fox or H-function
\cite{fox,mathai} which has the following alternating power series
expansion:
\begin{eqnarray}
H_{p,q}^{m,n}\left( z\; \vline
\begin{array}{c}
(a_j,A_j)_{j=1, \ldots ,p}\\ (b_j,B_j)_{j=1, \ldots ,q}
\end{array}
\right) & = & \sum_{l=1}^{m} \ \sum_{k=0}^{\infty} \ \frac{(-1)^k
z^{s_{lk}}}{k! B_l} \nonumber \\
 & & \times \frac{\prod_{j=1,j \neq l}^m \Gamma
 (b_j-B_j s_{lk}) \prod_{r=1}^n \Gamma(1-a_r+A_r
s_{lk})}{\prod_{u=m+1}^q \Gamma(1-b_u+B_u s_{lk}) \prod_{v=n+1}^p
\Gamma(a_v-A_v s_{lk})}, \label{hfn}
\end{eqnarray}
where $s_{lk} = (b_l+k)/B_l$ and an empty product is interpreted
as unity. Further, $m,n,p,q$ are nonnegative integers such that $0
\leq n \leq p$, $1 \leq m \leq q$; $A_j$, $B_j$ are positive
numbers; $a_j$, $b_j$ can be complex numbers. The H-function
has several remarkable properties \cite{mathai} which are listed in the Appendix.

By comparing the series expansion \cite{erdelyi} of the
Mittag-Leffler function $E_{\gamma}(z)$ with that of the
H-function [cf. Eq. (\ref{hfn})], Eq. (\ref{sol2}) can be
rewritten as
\begin{equation}
W(x,t) = \frac{2}{\pi} \int_0^{\infty}\  dk \sin kx \ \sin ka \
H_{1,2}^{1,1}\left( k^2 K t^{\gamma} \; \vline
\begin{array}{cc}
(0,1) & \\ (0,1), & (0,\gamma)
\end{array}
\right).
\end{equation}
Letting $k' = k (Kt^{\gamma})^{1/2}$ and using Property 5 [Eq.
(\ref{p4})] of H-functions, the above equation becomes
\begin{equation}
W(x,t) = \frac{1}{2 \pi (Kt^{\gamma})^{1/2}} \int_0^{\infty}\ dk'
[\cos k'(x-a) - \cos k'(x+a)] H_{1,2}^{1,1}\left( k' \; \vline
\begin{array}{cc}
(0,1/2) & \\ (0,1/2), & (0,\gamma/2)
\end{array}
\right).
\end{equation}
The Fourier cosine transforms can be solved by successive
applications of a Laplace and an inverse Laplace transform (a
technique pioneered by Fox \cite{fox2} for solving a wide variety
of integral transforms) to give \cite{srivastava}
\begin{eqnarray}
W(x,t) & = & \frac{1}{2 |x-a|} H_{3,3}^{2,1}\left(
\frac{|x-a|}{(Kt^{\gamma})^{1/2}} \; \vline
\begin{array}{ccc}
(1,1/2), & (1,\gamma/2), & (1,1/2)\\ (1,1), & (1,1/2), & (1,1/2)
\end{array}
\right) \nonumber \\ & &  - \frac{1}{2 (x+a)} H_{3,3}^{2,1}\left(
\frac{x+a}{(Kt^{\gamma})^{1/2}} \; \vline
\begin{array}{ccc}
(1,1/2), & (1,\gamma/2), & (1,1/2)\\ (1,1), & (1,1/2), & (1,1/2)
\end{array}
\right).
\end{eqnarray}
Now, applying Properties 2, 1, 3 and 6 of the H-functions (listed
in the Appendix) in the given order, we finally get
\begin{eqnarray}
&& W(x,t) = \nonumber \\
 && \frac{1}{2 (Kt^{\gamma})^{1/2}} \left[
H_{1,1}^{1,0}\left( \frac{|x-a|}{(Kt^{\gamma})^{1/2}} \; \vline
\begin{array}{c}
(1-\gamma/2,\gamma/2)\\ (0,1)
\end{array}
\right)- H_{1,1}^{1,0}\left( \frac{x+a}{(Kt^{\gamma})^{1/2}} \;
\vline
\begin{array}{c}
(1-\gamma/2,\gamma/2)\\ (0,1)
\end{array}
\right) \right]. \label{wkt}
\end{eqnarray}

Substituting Eq.~(\ref{wkt}) into Eq. (\ref{fT}) we have
\begin{eqnarray}
f(t) & = & -\frac{d}{dt} \left[ \frac{1}{2 (Kt^{\gamma})^{1/2}}
\int_0^{\infty} \ dx \ H_{1,1}^{1,0}\left(
\frac{|x-a|}{(Kt^{\gamma})^{1/2}} \; \vline
\begin{array}{c}
(1-\gamma/2,\gamma/2)\\ (0,1)
\end{array}
\right) \right] \nonumber \\ & &  +\frac{d}{dt} \left[ \frac{1}{2
(Kt^{\gamma})^{1/2}} \int_0^{\infty} \ dx \ H_{1,1}^{1,0}\left(
\frac{x+a}{(Kt^{\gamma})^{1/2}} \; \vline
\begin{array}{c}
(1-\gamma/2,\gamma/2)\\ (0,1)
\end{array}
\right) \right].
\end{eqnarray}
Removing the explicit time dependence in the integrands by rewriting
the integrals in terms of $z=(x-a)/(KT^{\gamma})^{1/2}$,
$z'=(x+a)/(KT^{\gamma})^{1/2}$, the above integrals can be
explicitly evaluated to give:
\begin{equation}
f(t) = \frac{a \gamma}{2 K^{1/2} t^{(2+\gamma)/2}}
H_{1,1}^{1,0}\left( \frac{a}{(Kt^{\gamma})^{1/2}} \; \vline
\begin{array}{c}
(1-\gamma/2,\gamma/2)\\ (0,1)
\end{array}
\right), \label{sol4}\end{equation} which is the main result of
this paper. It should be noted that H-functions were first used in
the context of probability distributions by Schneider
\cite{schneider}. They have also been used to express solutions of
fractional diffusion equations \cite{schneider2}. In addition, the
FPT problem in the context of Levy processes has been considered
in \cite{lindenberg}.

The series expansion of the H-function in Eq.~(\ref{sol4}) [cf.
Eq. (\ref{hfn})] is
\begin{equation}
f(t) = \frac{a \gamma}{2 K^{1/2} t^{(2+\gamma)/2}}
\sum_{k=0}^{\infty} \frac{(-a/(Kt^{\gamma})^{1/2})^k}{k!
\Gamma(1-\gamma/2-k \gamma/2)}.
\end{equation}
This turns out to be also the series expansion of the Maitland's
generalized hypergeometric function or the Wright function $_0
\psi_1$ \cite{erdelyi}. Thus, an alternative expression for $f(t)$
is
\begin{equation}
f(t) = \frac{a \gamma}{2 K^{1/2} t^{(2+\gamma)/2}} \ _0 \psi_1
\left(
\begin{array}{c}
- \\ (1-\gamma/2,-\gamma/2)
\end{array}; -\frac{a}{(KT^{\gamma})^{1/2}}
\right). \label{main}
\end{equation}
For $\gamma = 1$ (ordinary Brownian motion), the Wright function
reduces to the following simple formula
\begin{equation}
f(t) =  \frac{a}{(4 \pi K t^3)^{1/2}} e^{-a^2/4Kt}.
\end{equation}
This is the expected inverse Gaussian distribution for the FPT
distribution of the ordinary Brownian motion \cite{grimmet}.

Next, we consider the asymptotic behavior of the FPT distribution
for large values of $t$. Refer to Eq.~(\ref{sol4}). Let
$z=a/(Kt^{\gamma})^{1/2}$. It is known that
\cite{mathai,braaksma}, for small $z$, $H_{1,1}^{1,0}(z) \sim
|z|^{b_1/B_1}=1$, since $b_1=0$ and $B_1=1$. Therefore, the FPT
distribution $f(t)$, for large $t$, is characterized by the power
law relation
\begin{equation}
f(t) \sim t^{-1-\gamma/2}, \label{universal}
\end{equation}
which becomes the well known $-3/2$ scaling law for the ordinary
Brownian motion. This power law behavior has been observed earlier
by Balakrishnan \cite{balakrishnan} for subdiffusive processes ($0
< \gamma < 1$) using a different method. Using our method the same
scaling law is shown to be applicable also to superdiffusive
processes. After some manipulation, we can also determine the
location $t_{\rm max}$ of the maximum of the FPT distribution:
\begin{equation}
t_{\rm max} = \left( \frac{2 d \gamma}{4-\gamma}
\right)^{(2-\gamma)/\gamma},
\end{equation}
where $d=(1-\gamma/2)( \gamma/2 )^{\gamma/(2-\gamma)} ( a /
\sqrt{K} )^{2/(2-\gamma)}$. From Eq.~(\ref{universal}), we see
that the mean first passage time and all higher moments of the FPT
distribution are undefined for $0 < \gamma < 2$.

The theoretical prediction for the full FPT distribution given in
Eq. (\ref{sol4}) is verified by numerically simulating the
underlying CTRW process characterized by the probability density
function $\phi(y,u)$ [cf. Eq. (\ref{LWpdf})]. For the sake of
numerical efficiency, we replace the waiting time distribution in
$\phi(y,u)$ by the Pareto distribution \cite{evans} which is well
justified for small values of $\tau$. Ten million realizations of
the CTRW process are used to generate the numerical FPT
distribution. The results are shown in Figure 1 for $\gamma =
0.5$, $a=1.0$, $\tau=10^{-4}$ and $K=0.1$. We note that the
numerical simulation is in excellent agreement with the
theoretical prediction. The agreement would get even better as the
generalized diffusion limit is approached (that is, as $\tau \to
0$ and $\sigma^2 \to 0$ with $K$ held a constant).

\section{Anomalous Diffusion of the fBm Type}

Fractional Brownian motion $X(t)$ \cite{mandelbrot,taqqu} is a
Gaussian process with $X(0)=0$, $<X(t)> = 0$ and $< [X(t)-X(s)]^2
> = |t-s|^{\gamma}$ ($0 < \gamma < 2$). By definition it provides
us with another type of anomalous diffusion. The exact FPT
distribution of this process is not known. It was conjectured
\cite{ding}, based on scaling argument and numerical evidence,
that for large $t$, $f(t)$ scales with $t$ as
\begin{equation}\label{fbmscal}
f(t) \sim t^{\gamma/2-2}.
\end{equation}
Notice that this power law behavior is different from that in
Eq.~(\ref{universal}) even though the mean square displacement $<
X(t)^2>$ has the same power law behavior ($< X(t)^2> \sim
t^{\gamma}$) for both types of anomalous diffusion. Below we give
a heuristic argument for this power law using a recent result
\cite{molchan} concerning the distribution of the maximum of a fBm
over a given interval.

Without loss of generality we set the threshold at $a=1$. Let the
probability that the maximum $M_t$ of the fBm $X(t)$ with $X(0)=0$
is less than 1 in the time interval $[0,t]$ be denoted by $P(t)$:
\begin{equation}
P(t) = {\rm Prob}(M_t < 1)
\end{equation}
Clearly, $P(t)$ is also the probability that the first passage
time $T$ of the fBm is greater than $t$:
\begin{equation}
{\rm Prob}(T > t) = P(t).
\end{equation}
This implies that the first passage time distribution $f(t)$ is
given by
\begin{equation}\label{max}
f(t) = \frac{d}{dt} {\rm Prob}(T \le t) = - \frac{d}{dt} {\rm
Prob}(T \ge t) = - \frac{d}{dt} P(t).
\end{equation}
Recent work by Molchan \cite{molchan} shows that, in the large $t$
limit, $P(t)$ scales with $t$ as
\begin{equation}
P(t) \sim t^{\gamma/2-1}, \ \ \ t \to \infty.
\end{equation}
Substituting this in Eq. (\ref{max}) we obtain
Eq.~(\ref{fbmscal}).

We present numerical results to verify Eq.~(\ref{fbmscal}).
Sinai's formula \cite{sinai} for the power spectrum of the
fractional Gaussian noise (fGn) is used to generate the fBm. The
log-log plot of the FPT distribution is shown in Figure 2
($\gamma=1.5$). It is clear that the predicted slope of
$\gamma/2-2=-1.25$ is in excellent agreement with the numerical
simulation.

\section{Conclusions}

The two types of anomalous diffusions considered in this work lead
to two distinct scaling behavior, Eq.~(\ref{universal}) and
Eq.~(\ref{fbmscal}), for the respective FPT distributions in the
asymptotic limit, despite the fact that they are both described by
the same mean square displacement. Equation (\ref{universal}) is
expected to be applicable to all CTRW types of processes,
regardless of the specific forms of $\phi(y,u)$, for which the
generalized diffusion limit leads to Eq.~(\ref{FFPE}). On the
other hand, we expect Eq.~(\ref{fbmscal}) to hold for Gaussian
processes where $<[X(t)-X(s)]^2> \sim |t-s|^{\gamma}$ for large
$|t-s|$.

In this work we considered only processes with $<X(t)>=0$ where
the asymptotic limits of the FPT distributions are described by
power laws. For a fBm, little is known about its FPT distribution
when $<X(t)> \neq 0$. For a Levy type of diffusion process, some
exact results can be derived for the Laplace transform of the FPT
distribution when $<X(t)> \neq 0$. We will present these results
in other publications.

\section{Acknowledgements}

The work was supported by US ONR Grant N00014-99-1-0062. GR thanks
Center for Complex Systems and Brain Sciences, Florida Atlantic
University, where this work was performed, for hospitality. He is
also associated with the Jawaharlal Nehru Center for Advanced
Scientific Research as a honorary faculty member. The authors
would like to thank Mike Shlesinger and Kiran Kolwankar for useful
discussions. They would also like to thank the referees for their
valuable comments.

\newpage
\appendix
\section*{Properties of H-functions}

The H-function has the following properties \cite{mathai} which
have used in the main text.

{\em Property 1.} The H-function is symmetric in the pairs
$(a_1,A_1), \ldots ,(a_n,A_n)$, likewise $(a_{n+1},A_{n+1}),
\ldots ,(a_p,A_p)$; in $(b_1,B_1), \ldots ,(b_m,B_m)$ and in
$(b_{m+1},B_{m+1}), \ldots ,(b_q,B_q)$.

{\em Property 2.} Provided $n \geq 1$ and $q > m$,
\begin{eqnarray} \label{p2}
&& H_{p,q}^{m,n}\left( z \; \vline
\begin{array}{cccc}
(a_1,A_1), & (a_2,A_2), & \cdots , & (a_p,A_p)\\
(b_1,B_1), & \cdots , & (b_{q-1},B_{q-1}), & (a_1,A_1)
\end{array}
\right) \nonumber \\ & = & H_{p-1,q-1}^{m,n-1}\left( z \; \vline
\begin{array}{cccc}
(a_2,A_2), & \cdots , & (a_p,A_p)\\
(b_1,B_1), & \cdots , & (b_{q-1},B_{q-1})
\end{array}
\right).
\end{eqnarray}

{\em Property 3.} Provided $m \geq 2$ and $p > n$,
\begin{eqnarray}\label{p3}
&& H_{p,q}^{m,n}\left( z \; \vline
\begin{array}{cccc}
(a_1,A_1), & \cdots , & (a_{p-1},A_{p-1}), & (b_1,B_1)\\
(b_1,B_1), & (b_2,B_2), & \cdots , & (b_q,B_q)
\end{array}
\right) \nonumber \\
 & = & H_{p-1,q-1}^{m-1,n}\left( z \; \vline
\begin{array}{cccc}
(a_1,A_1), & \cdots , & (a_{p-1},A_{p-1})\\
(b_2,B_2), & \cdots , & (b_q,B_q)
\end{array}
\right).
\end{eqnarray}

{\em Property 4.}
\begin{equation}\label{p4}
H_{p,q}^{m,n}\left( z \; \vline
\begin{array}{c}
(a_j,A_j)_{j=1, \ldots ,p}\\
(b_j,B_j)_{j=1, \ldots ,q}
\end{array}
\right) = H_{q,p}^{n,m}\left( \frac{1}{z} \; \vline
\begin{array}{c}
(1-b_j,B_j)_{j=1, \ldots ,q} \\
(1-a_j,A_j)_{j=1, \ldots ,p}
\end{array}
\right).
\end{equation}

{\em Property 5.} For $k > 0$,
\begin{equation}\label{p5}
\frac{1}{k} H_{p,q}^{m,n}\left( z \; \vline
\begin{array}{c}
(a_j,A_j)_{j=1, \ldots ,p}\\
(b_j,B_j)_{j=1, \ldots ,q}
\end{array}
\right) = H_{p,q}^{m,n}\left( z^k \; \vline
\begin{array}{c}
(a_j,k A_j)_{j=1, \ldots ,p}\\
(b_j,k B_j)_{j=1, \ldots ,q}
\end{array}
\right).
\end{equation}

{\em Property 6.}
\begin{equation}\label{p6}
z^{\rho} H_{p,q}^{m,n}\left( z \; \vline
\begin{array}{c}
(a_j,A_j)_{j=1, \ldots ,p}\\
(b_j,B_j)_{j=1, \ldots ,q}
\end{array}
\right) = H_{p,q}^{m,n}\left( z \; \vline
\begin{array}{c}
(a_j+\rho A_j,A_j)_{j=1, \ldots ,p}\\
(b_j+\rho B_j,B_j)_{j=1, \ldots ,q}
\end{array}
\right).
\end{equation}

\newpage

\section*{Figure Legends}

\begin{description}

\item{\bf Figure 1:} Comparison of the theoretical FPT
distribution (solid line) with the distribution (dashed line)
obtained by numerically simulating the underlying CTRW process for
a Levy type anomalous diffusion with $\gamma=0.5$.

\item{\bf Figure 2:} Comparison of the theoretically predicted
power law behavior of the FPT distribution for a fBm with
$\gamma=1.5$ (dashed line) with the numerical simulation (solid
line).

\end{description}

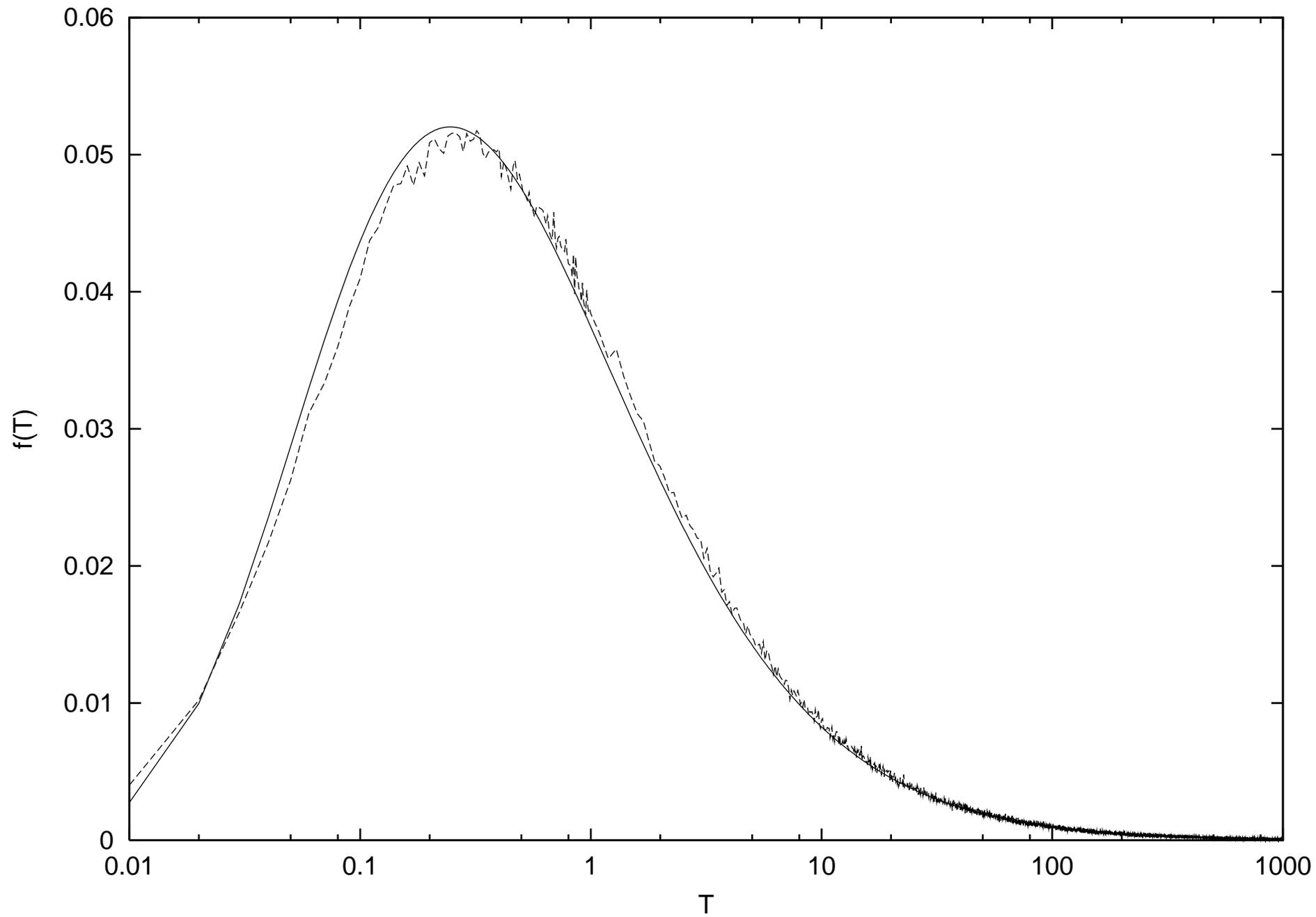
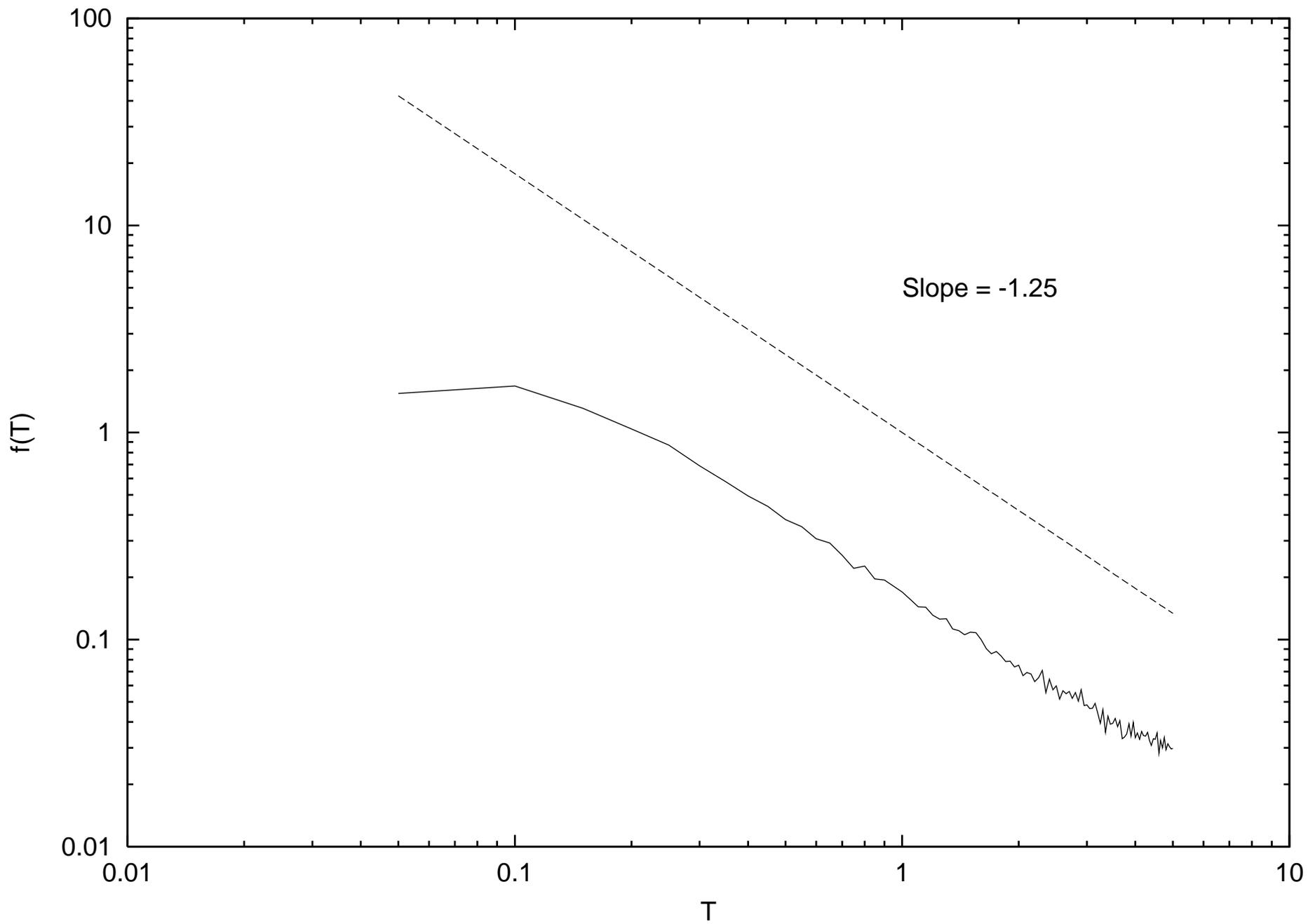
\end{document}